\documentclass[pre,twocolumn,superscriptaddress]{revtex4}
\usepackage{amsmath}
\usepackage{amssymb}

\usepackage{amsmath,amssymb}
\usepackage[usenames]{color}
\usepackage{amssymb}
\usepackage{grffile}
\usepackage[pdftex]{graphicx}
\usepackage{amsmath, amstext, amssymb, amsfonts, amsxtra}
\usepackage{textcomp}
\usepackage{xspace} 
\usepackage[colorlinks]{hyperref}
\usepackage{tikz}
\usetikzlibrary{shapes,arrows,fit,positioning}

\newcommand{\Hop}{\hat{H}} 

\newcommand{\aop}{\hat{a}}
\newcommand{\adop}{\hat{a}^{\dagger}}
\newcommand{\nop}{\hat{n}}

\newcommand{\rhoop}{\hat{\rho}}

\newcommand{\sgx}{\hat{\sigma}^x}
\newcommand{\sgy}{\hat{\sigma}^y}
\newcommand{\sgz}{\hat{\sigma}^z}
\newcommand{\sgp}{\hat{\sigma}^+}
\newcommand{\sgm}{\hat{\sigma}^-}

\newcommand{\hc}{{\rm H.c.}} 
\newcommand{\im}{{\rm i}} 
\newcommand{\tr}{{\rm tr}}

\newcommand{\id}{{\textbf{1}}} 

\newcommand{\conj}{{\rm conjugate}}

\newcommand{\svd}{{\rm SVD}}

\newcommand{\SYM}[1]{\mathcal{#1}}
\newcommand{\BD}[1]{\mathbf{#1}}
\newcommand{\Heff}{{{\hat{\rm H}_{\rm eff}}}}
\newcommand{\Hleft}{\SYM{L}}
\newcommand{\Hright}{\SYM{R}}
\newcommand{\Ham}{\hat{{\rm H}}}

\newcommand{\Evo}{\hat{{\rm U}}}

\newcommand{\dt}{{\rm dt}}
\newcommand{\MPS}{{\rm MPS}}
\newcommand{\MPO}{{\rm MPO}}
\newcommand{\SMPS}{{\rm s}\text{-}{\rm MPS}}
\newcommand{\DSMPS}{{\rm as}\text{-}{\rm MPS}}
\newcommand{\SMPO}{{\rm s}\text{-}{\rm MPO}}
\newcommand{\DSMPO}{{\rm as}\text{-}{\rm MPO}}
\newcommand{\XXZ}{{\rm XXZ}}
\newcommand{\XYZ}{{\rm XYZ}}

\newcommand{\zero}{\textbf{0}} 

\makeatletter
\DeclareRobustCommand{\cev}[1]{%
  {\mathpalette\do@cev{#1}}%
}
\newcommand{\do@cev}[2]{%
  \vbox{\offinterlineskip
    \sbox\z@{$\m@th#1 x$}%
    \ialign{##\cr
      \hidewidth\reflectbox{$\m@th#1\vec{}\mkern4mu$}\hidewidth\cr
      \noalign{\kern-\ht\z@}
      $\m@th#1#2$\cr
    }%
  }%
}
\makeatother

\newcommand{\Lop}{\hat{\mathcal{L}}}
\newcommand{\Dop}{\hat{\mathcal{D}}}

\newcommand{\sutd}{Science and Math Cluster and EPD, Singapore University of Technology and Design, 8 Somapah Road, 487372 Singapore} 

\newcommand{\sft}{Quantum Intelligence Lab (QI-Lab), Supremacy Future Technologies (SFT), Guangzhou 511340, China}

\begin{document}

\title{Matrix Product States with adaptive global symmetries}
\author{Chu Guo}
\affiliation{\sft}
\author{Dario Poletti} 
\affiliation{\sutd}

\begin{abstract}    
Quantum many body physics simulations with Matrix Product States can often be accelerated if the quantum symmetries present in the system are explicitly taken into account. Conventionally, quantum symmetries have to be determined before hand when constructing the tensors for the Matrix Product States algorithm. In this work, we present a Matrix Product States algorithm with an adaptive $U(1)$ symmetry. This algorithm can take into account of, or benefit from, $U(1)$ or $Z_2$ symmetries when they are present, or analyze the non-symmetric scenario when the symmetries are broken without any external alteration of the code. To give some concrete examples we consider an XYZ model and show the insight that can be gained by (i) searching the ground state and (ii) evolving in time after a symmetry-changing quench. To show the generality of the method, we also consider an interacting bosonic system under the effect of a symmetry-breaking dissipation.     
\end{abstract}

\date{\today}
\maketitle

\address{} 

\vspace{8mm}

\section{introduction}
Matrix Product States (MPS) algorithms have become one of the best tools to study one-dimensional quantum manybody systems \cite{FannesWerner1992, OstlundRommer1995, Perez-GarciaCirac2007, Schollwock2011}. It has been shown that they can represent the ground state of a wide class of local Hamiltonians \cite{Hastings2007, Vidal2008}. They can also be applied to study the time evolution of both unitary and open quantum manybody systems \cite{Vidal2003, Vidal2004, DaleyVidal2004, WhiteFeiguin2004, VerstraeteCirac2004, Daley2014, DeVegaBanuls2015, GuoPoletti2018a, XuPoletti2019, ProsenZnidaric2009, PalmeroPoletti2019}, provided the entanglement entropy, or the operator space entanglement entropy for dissipative systems \cite{PizornProsen2008}, do not grow too fast.

Symmetries play a very important role in physics as they imply the presence of conserved quantities. These make it possible to write the Hamiltonian of a system in a block-diagonal form, where each block has an associated quantum number, thus significantly reducing the storage requirements and the computational cost of diagonalizing it. 
In a number of systems and models the total number of particles, or the total magnetization, are conserved: in these scenarios the systems have a $U(1)$ symmetry. In other cases a system could have a discrete symmetry, like a $Z_2$ symmetry, which is a subgroup of $U(1)$. In such situations, it is often advantageous to explicitly encode the quantum symmetry into the tensors of the MPS \cite{DaleyVidal2004, Perez-GarciaVidal2008, SinghVidal2011}, which are also called symmetric Matrix Product States ($\SMPS$). $\SMPS$ would usually result in more efficient numerical algorithms because it only explores the portion of the Hilbert space corresponding to the relevant symmetry sector. $\SMPS$ are also particularly important when searching for the ground state in a certain symmetry sector of an Hamiltonian, otherwise the ground state search algorithm would return the lowest ground state of the global Hilbert space, i.e. of all the symmetry sectors. Typically, one would thus use a different MPS depending on whether the Hamiltonian is symmetric or not, or maybe use a lesser efficient option. This is particularly true in the study of dynamical systems in which the equations of motion change the symmetry properties of the system during the evolution.             
%
%
%
For example, one could have prepared a quantum state in the ground state of an $\XXZ$ chain, which has $U(1)$ symmetry, and then suddenly tune the parameter so that the system is modeled by a $Z_2-$symmetric $\XYZ$ chain. With conventional $\SMPS$, one would have to manually convert the ground state from $U(1)$ to $Z_2$ symmetry for the time evolution.

In this work we propose MPS with adaptive global $U(1)$ symmetry to overcome this difficulty, which we will refer to as adaptively symmetric Matrix Product States ($\DSMPS$). The central idea of our approach is to generalize $U(1)$ $\SMPS$ so that instead of a fixed total quantum number, it allows a superposition of different total quantum numbers only. Importantly, the amount of different quantum numbers required changes automatically during the calculations including only the quantum numbers required. As a result, a $U(1)$ $\DSMPS$ is able to treat systems with $U(1)$ symmetry or systems whose symmetry group is a subgroup of $U(1)$, or even no symmetry at all, on the same footing. 

This paper is divided into the following sections: In Sec.\ref{sec:method}, we introduce in detail how to implement $U(1)$ $\DSMPS$, including the ground state search and time evolution algorithms; In Sec.\ref{sec:app}, we show two concrete examples in which we can demonstrate the use of $\DSMPS$, one for a unitary system and the other for a dissipative system; In Sec.\ref{sec:summary} we draw our conclusions.

\section{method}\label{sec:method}
Similarly to $\SMPS$, the building blocks of $\DSMPS$ are symmetry protected tensors. The definition of a $U(1)$ symmetry protected tensor, as well as the basic tensor operations based on it, has been presented in detail in the literature. One particularly detailed example is \cite{ SinghVidal2011}, although the use of symmetry protected tensors has been presented earlier, for instance in \cite{McCulloch2007}. In Sec.\ref{ssec:spt}, we will first give a minimal introduction to symmetry protected tensors, then in Secs.\ref{ssec:mps} and \ref{ssec:mpo} we, respectively, describe the adaptively symmetric MPS and Matrix Product Operator (MPO). In Sec.\ref{ssec:groundstate} we describe how to search for the ground state and in Sec.\ref{ssec:timeevolution} how to implement time evolution.  

\subsection{Symmetry protected tensors}\label{ssec:spt} 
A $U(1)$ symmetry protected tensor can be represented as a list of dense tensors labelled by quantum numbers, the dimension of the space corresponding to that quantum number and, very importantly, a direction for each quantum number stating whether it is incoming or outgoing from the tensor. This direction is key to preserve the symmetry for the overall state. For example, a $4$ dimensional symmetry protected tensor $\SYM{O}$ could be written as 
\begin{align}
\SYM{O} = \{\BD{O}^{(\vec{n}_1, I_{n_1}), (\vec{n}_2, I_{n_2})}_{(\cev{n}_3, I_{n_3}), (\cev{n}_4, I_{n_4})}\}. \label{eq:ex__spt}
\end{align}
Here $\BD{O}^{(\vec{n}_1, I_{n_1}), (\vec{n}_2, I_{n_2})}_{(\cev{n}_3, I_{n_3}), (\cev{n}_4, I_{n_4})}$ is a $4$ dimensional dense tensor with dimensions $I_{n_i}$, and $n_i$ for $1\leq i\leq 4$ labels the four quantum number (note that here we don't differentiate between superscripts and subscripts). The quantum numbers have two directions, $\vec{n}_i$ and $\cev{n}_i$, depending on whether, respectively, the quantum numbers flows in or out. The $U(1)$ symmetry can be explicitly enforced by the fusion rule
\begin{align} \label{eq:fusionrule}
n_1 + n_2 - n_3 - n_4 = 0,  
\end{align}
where $n_3$ and $n_4$ are subtracted because their quantum numbers flow outside of the tensor $\SYM{O}$, while $n_1$ and $n_2$ are added because their quantum numbers flow inside of the tensor $\SYM{O}$. For the implementation of algorithms with symmetric Matrix Product States, it is convenient to use an appropriate library that takes into account which tensors can be added, multiplied, decomposed and merged. Of particular importance would be a structured tensor which takes into account of the possible number conserving combinations (see, for example, \cite{SinghVidal2011}).           
To simplify the notation, in the following we will denote $(\vec{n}_i, I_{n_i})$ simply as $\vec{n}_i$ and $(\cev{n}_i, I_{n_i})$ as $\cev{n}_i$, implicitly indicating the size $I_{n_i}$; the tensor in Eq.(\ref{eq:ex__spt}) would then be rewritten as $\SYM{O} = \{\BD{O}^{\vec{n}_1, \vec{n}_2}_{\cev{n}_3, \cev{n}_4}\}$. 

\subsection{Adaptively symmetric Matrix Product States}\label{ssec:mps}

\begin{figure}
\includegraphics[width=\columnwidth]{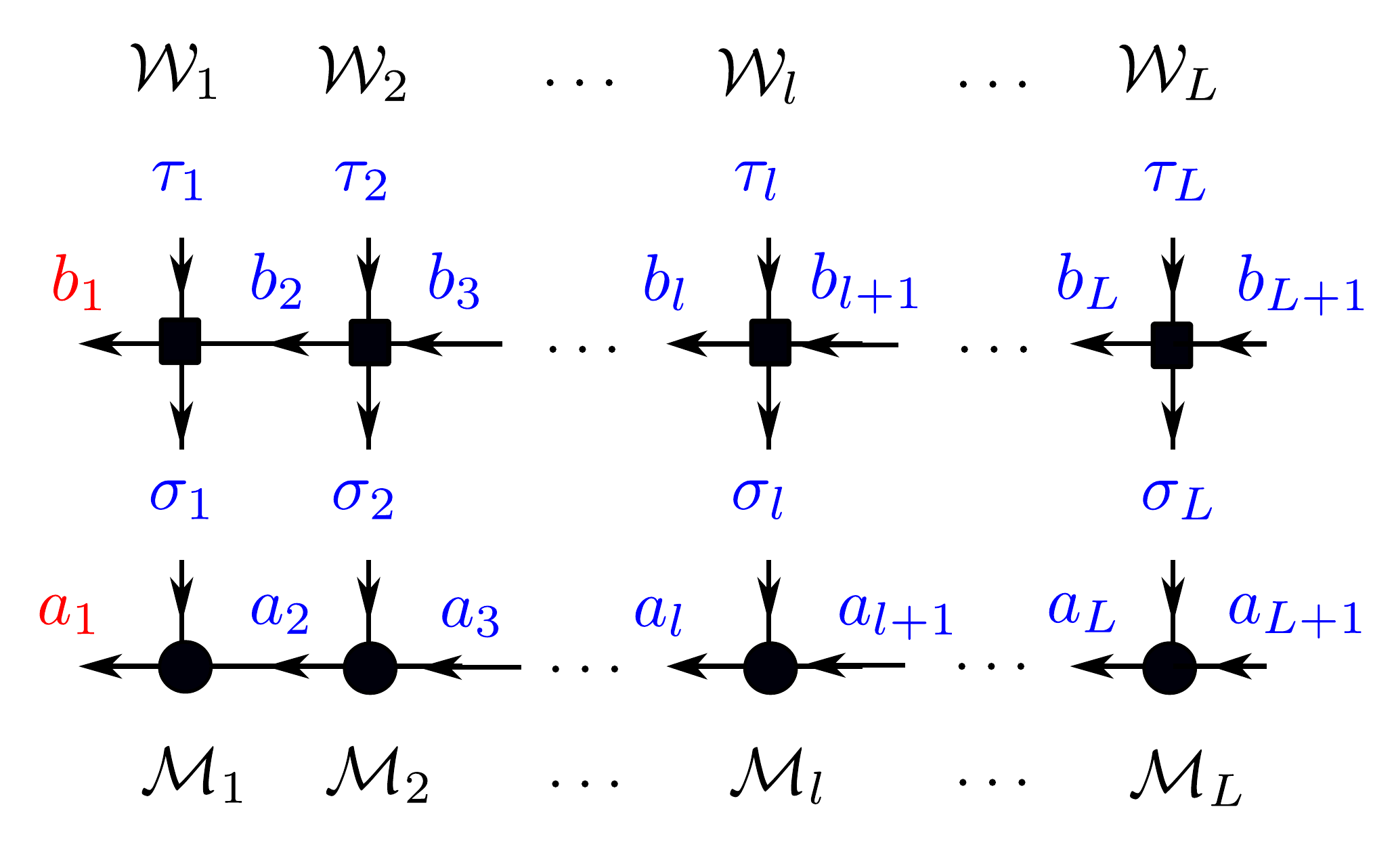}
\caption{$\DSMPS$ and $\DSMPO$ with adaptive global $U(1)$ symmetry. We use the convention that the quantum numbers flow in from the right hand side and out from the left hand side. Each $\mathcal{M}_l$ is a $3$ dimensional tensor and each $\mathcal{W}_l$ is a $4$ dimensional tensor. $a_{1}$ and $a_{L+1}$ are the quantum numbers on the left and right boundaries of the $\DSMPS$, and $b_{1}$ and $b_{L+1}$ are the quantum numbers on the left and right boundaries of the $\DSMPO$. $a_{L+1}$ and $b_{L+1}$ are usually chosen so that $a_{L+1}=b_{L+1}=0$. For $\SMPS$ and $\SMPO$, one has $a_1=N$ ($N$ is the total quantum number) and $b_1=0$ (the MPO conserves the total quantum number). For $\DSMPS$, $a_1$ is allowed to have multiple values $N_1, N_2, \dots$, each of them corresponds to an $\SMPS$ with total quantum number $N_1, N_2, \dots$. For $\DSMPO$, $b_1$ is allowed to have multiple values such as $-1, 0, 1, \dots$, each of them corresponding to a quantum operator which changes the total quantum number by $-1, 0, 1, \dots$ when applied to an $\DSMPS$. The labels $a_1$ and $b_1$ are in red to indicate the main difference between $\SMPS$ and $\SMPO$ with $\DSMPS$ and $\DSMPO$.}   \label{fig:fig1} 
\end{figure}

An MPS representing a quantum state on $L$ sites is a chain with $3-$dimensional tensors. The tensor $\SYM{M}_l$ on site $l$ of the MPS is labelled by a physical quantum number $\sigma_l$, an auxiliary quantum number $a_l$ which connects it to its left hand site, and $a_{l+1}$ which connects it to its right hand site. As a result, $\SYM{M}_{l}$ can be written as
\begin{align}
\SYM{M}_l = \{\BD{M}^{\cev{\sigma}_l}_{\vec{a}_l, \cev{a}_{l+1}}\},
\end{align}
and the fusion rule Eq.(\ref{eq:fusionrule}) becomes
\begin{align}
\sigma_l + a_{l+1} - a_l = 0.
\end{align}

We encode the flow of quantum numbers as in Fig.\ref{fig:fig1} (bottom). Conventionally, if an MPS has a fixed total quantum number $N$, then one would have $a_{L+1}=0$ for the right boundary, and $a_1=N$ for the left boundary, and we could denote it as $\SMPS_N$. The central observation is that one can generalize this approach such that $a_{L+1}=0$ still holds, while $a_1$ could have multiple values depending on how many possible total quantum numbers the quantum state has. As a result, different $a_1$ will correspond to a different total quantum number. Simply speaking, an $\DSMPS$ is a collection of the relevant $\SMPS$s which we write symbolically as 
\begin{align}
\DSMPS = \sum_{N} \SMPS_N.
\end{align}
As for typical $\SMPS$, the {\it overall} bond dimension (defined as the size of the auxiliary dimensions) of an $\DSMPS$ is the sum of the sizes of all the blocks corresponding to different auxiliary quantum numbers. In this way, while keeping each block to a manageable size, it is possible to effectively use a very large overall bond dimension for the $\DSMPS$.

\subsection{Adaptively symmetric Matrix Product Operators}\label{ssec:mpo}
Adaptively symmetric Matrix Product Operators ($\DSMPO$) can be treated in a similar manner as they form a chain of $4-$dimensional symmetry protected tensors. The tensor $\SYM{W}_l$ on the $l$-th site of an $\DSMPO$ is labelled by two physical quantum numbers $\sigma_l$ and $\tau_l$, an auxiliary quantum number $b_l$ which connects it to its left hand site, and $b_{l+1}$ which connects it to its right hand site. Thus $\SYM{W}_l$ can be written as
\begin{align}
\SYM{W}_l = \{\BD{W}^{\vec{\sigma}_l, \cev{\tau}_l}_{\vec{b}_l, \cev{b}_{l+1}}\},
\end{align}
where the dimensions of the tensors are implicit in this notation, and the quantum numbers satisfy
\begin{align}
\tau_l + b_{l+1} - \sigma_l - b_l = 0.
\end{align} 
The flow of quantum numbers is represented in Fig.\ref{fig:fig1} (top). For $U(1)$ symmetric systems, the MPO does not change the total quantum number of the quantum state. If the quantum state had total quantum number $N$ before being multiplied by the MPO, after the multiplication between MPO and MPS, the total quantum number for the quantum state is still $N$.  As a result we can denote a symmetric MPO as $\SMPO=\sum_N\SMPO_N^{N}$, with the particular number preserving condition that ${b_1}={b_{L+1}}=0$. 
In order to generalize this concept, we allow the quantum number $b_1$ at the left boundary to take multiple values while still keeping $b_{L+1}=0$. 
Therefore the adaptively symmetric Matrix Product Operator can be written as
\begin{align} \label{eq:splitmps}
\DSMPO = \sum_{N, N'} \SMPO_N^{N'},
\end{align}
where $N'=N$ if and only if the system has $U(1)$ symmetry. In case the system has only a discrete symmetry, for example parity, a $Z_2$ symmetry, then $N$ and $N'$ satisfy $N'=N\pm E$ where $E$ is an even number. If the system has no symmetry at all, then $N$ and $N'$ can be arbitrary. However, in this case, it may be more efficient to use non-symmetric $\MPO$ and $\MPS$ since the number of states will be the same for both approaches and the non-symmetric version has a much simpler data structure.

In the next sections we show how to adapt the ground state search and time evolution algorithms using $\DSMPS$ and $\DSMPO$.

\subsection{Ground state search algorithm} \label{ssec:groundstate}
Here we present the ground state search algorithm that we have implemented based on $\DSMPS$. The core structure retraces that of a two-site variational Matrix Product States algorithm which we briefly present in the following. In this algorithm it is convenient to keep the MPS in the ``mixed-canonical'' form
\begin{align}\label{eq:mixedcanonical}
\sum_{a_1, a_2, \dots, a_{L+1}} \SYM{A}_1\dots\SYM{A}_{l-1}\SYM{M}_l \SYM{M}_{l+1}\SYM{B}_{l+2}\dots\SYM{B}_L,
\end{align}
where the $\SYM{A}_l$ are called ``left-canonical'' because 
\begin{align}\label{eq:leftcanonicalcon}
\sum_{\sigma_l, a_{l}}\SYM{A}^{\cev{\sigma}_l}_{\vec{a}_l, \cev{a}_{l+1}} \conj(\SYM{A})^{\vec{\sigma}_l}_{\cev{a}_l, \vec{a}'_{l+1}} = \SYM{I}_{\cev{a}_{l+1}, \vec{a}_{l+1}'},
\end{align}
and the $\SYM{B}_{l}$ are ``right-canonical'', i.e. 
\begin{align}\label{eq:rightcanonicalcon}
\sum_{\sigma_l, a_{l+1}}\SYM{B}^{\cev{\sigma}_l}_{\vec{a}_l, \cev{a}_{l+1}} \conj(\SYM{B})^{\vec{\sigma}_l}_{\cev{a}_l', \vec{a}_{l+1}} = \SYM{I}_{\vec{a}_l, \cev{a}_l'}.
\end{align}
In both Eqs.(\ref{eq:leftcanonicalcon},\ref{eq:rightcanonicalcon}) $\SYM{I}_{\vec{a}_l, \cev{a}_l'}$ represents the identity matrix for the two indices $\vec{a}_l$, $\cev{a}_l'$.      
Here $\conj(\SYM{O})$ means to take the element wise conjugate of the tensor $\SYM{O}$, during which the direction of each quantum number also gets reversed. The two tenors with a commom quantum number $n$ which is summed over should have opposite directions. No restrictions is made on the tensors $\SYM{M}$. A two-site ground state search variational MPS algorithm works by iteratively minimizing the energy of an effective Hamiltonian $\Heff$ on each neighbouring pair of sites. This minimization is done iteratively, from left to right and then back from right to left, in what is called a sweep. Thus the central step for the two-site ground state search algorithm is to build the local effective Hamiltonian $\Heff$. To compute $\Heff$ on sites $l$ and $l+1$ (also referred to as bond $l$), one needs to trace out the \textit{left environment} consisting of the sites to the left of the $l$-th site to obtain the tensor $\Hleft_{l}$
\begin{align}
&\Hleft_{\vec{a}_l', \cev{b}_l, \cev{a}_l} 
= \sum_{\substack{{a_{l-1}', b_{l-1}, a_{l-1}}\\
    {\sigma_{l-1}, \sigma_{l-1}'}}} \Hleft_{\vec{a}_{l-1}', \cev{b}_{l-1}, \cev{a}_{l-1}}  \;\; \times \nonumber \\ 
&\SYM{A}_{\vec{a}_{l-1}, \cev{a}_l}^{\cev{\sigma}_{l-1}}  \;\; \conj(\SYM{A})_{\cev{a}_{l-1}', \vec{a}_l'}^{\vec{\sigma}_{l-1}'} \SYM{W}_{\vec{b}_{l-1}, \cev{b}_l}^{\vec{\sigma}_{l-1}, \cev{\sigma}_{l-1}'},
\end{align}
and the \textit{right environment} consisting of sites to the right of the $\left(l+1\right)$-th site to obtain the tensor $\Hright_{l+2}$
\begin{align}
&\Hright_{\cev{a}_{l+2}', \vec{b}_{l+2}, \vec{a}_{l+2}} = \sum_{\substack{{a_{l+3}', b_{l+3}, a_{l+3}}    \\
    {\sigma_{l+2}, \sigma_{l+2}'}}}\Hright_{\cev{a}_{l+3}', \vec{b}_{l+3}, \vec{a}_{l+3}} \;\; \times \nonumber \\ 
&\SYM{B}_{\vec{a}_{l+2}, \cev{a}_{l+3}}^{\cev{\sigma}_{l+2}} \conj(\SYM{B})_{\cev{a}_{l+2}', \vec{a}_{l+3}'}^{\vec{\sigma}_{l+2}'} \SYM{W}_{\vec{b}_{l+2}, \cev{b}_{l+3}}^{\vec{\sigma}_{l+2}, \cev{\sigma}_{l+2}'}.
\end{align}
With $\Hleft_l$, $\Hright_{l+2}$ and $\SYM{W}_l$, $\SYM{W}_{l+1}$, one can construct $\Heff$ as
\begin{align}\label{eq:heff}
\hat{\rm H}_{{\rm eff}\; \cev{a}_l, \vec{\sigma}_l, \vec{\sigma}_{l+1}, \vec{a}_{l+2}}^{\;\;\;\;\;\vec{a}_l', \cev{\sigma}_l', \cev{\sigma}_{l+1}', \cev{a}_{l+2}'} =& \sum_{b_l, b_{l+1} b_{l+2}} \Hleft_{\vec{a}_l', \cev{b}_l, \cev{a}_l} \Hright_{\cev{a}_{l+2}', \vec{b}_{l+2}, \vec{a}_{l+2}} \; \times \nonumber \\ 
&\SYM{W}_{\vec{b}_{l}, \cev{b}_{l+1}}^{\vec{\sigma}_l, \cev{\sigma}_l'} \SYM{W}_{\vec{b}_{l+1}, \cev{b}_{l+2}}^{\vec{\sigma}_{l+1}, \cev{\sigma}_{l+1}'}
\end{align}
Grouping the tensor indexes $(\cev{a}_l, \vec{\sigma}_l, \vec{\sigma}_{l+1}, \vec{a}_{l+2})$ and $(\vec{a}_l', \cev{\sigma}_l', \cev{\sigma}_{l+1}', \cev{a}_{l+2}')$, one could treat $\Heff$ as a matrix and find its lowest eigenstate $\SYM{M}_{\vec{a}_l, \cev{\sigma}_l, \cev{\sigma}_{l+1}, \cev{a}_{l+2}}$ and eigenvalue. 
$\SYM{M}_{\cev{a}_l, \vec{\sigma}_l, \vec{\sigma}_{l+1}, \vec{a}_{l+2}}$ is then decomposed into two $3$ dimensional tensors by singular value decomposition (SVD)
\begin{align}
\svd(\SYM{M}_{\vec{a}_l, \cev{\sigma}_l, \cev{\sigma}_{l+1}, \cev{a}_{l+2}}) = \sum_{s}\SYM{A}_{\vec{a}_l, \cev{s}}^{\cev{\sigma}_l} \; \SYM{S}_{\vec{s}, \cev{s}} \; \SYM{B}_{\vec{s}, \cev{a}_{l+2}}^{\cev{\sigma}_{l+1}},
\end{align}
and the results are used to update $\SYM{A}_l$ or $\SYM{B}_{l+1}$ depending on whether the sweep is from left to right or from right to left. 

To adapt the above algorithm to $\DSMPS$, one notices that the only difference between $\SMPS$, $\SMPO$ and $\DSMPS$, $\DSMPO$ is that for the latter there is a much larger variety of possible values for the left-most index ($a_1$ or $b_1$) of the tensors at the left boundary. Conventionally, ${a_1}$ (auxiliary index on the $\SMPS$) can only take a single value $N$ and ${b_1}$ (auxiliary index on the $\SMPO$) has to be $0$, therefore one can straightforwardly construct $\mathcal{L}_1$ as
\begin{align} 
\Hleft_{1}^{\SMPS} = \{ \textbf{1}_{\vec{N}, \cev{0}, \cev{N}}\},
\end{align}
where $\textbf{1}$ means a trivial dense tensor with a single element $1$ and where we have used the label $\SMPS$ to remind the reader that this concerns symmetric Matrix Product States. For $\DSMPS$ and $\DSMPO$, instead, $\Hleft_1$ must be able to include all possible quantum numbers that could appear after the $\DSMPO$ has been applied on the $\DSMPS$. Thus a way to construct $\Hleft_1$ is to look at the boundary indexes ${a_1}$ and ${b_1}$ and find all the relevant combinations $a_1+b_1$, namely
\begin{align} \label{eq:Hleft1}
\Hleft_1 = \{\textbf{1}_{\vec{a}_1+\vec{b}_1, \cev{b}_1, \cev{a}_1}\}.
\end{align}
With such a construction of $\Hleft_1$, the variational ground state search algorithm would be able to automatically detect and preserve the symmetry in the system, whether it is $U(1)$ or, for example, $Z_2$.

\subsection{Time evolution algorithm} \label{ssec:timeevolution}
\begin{figure}
\includegraphics[width=\columnwidth]{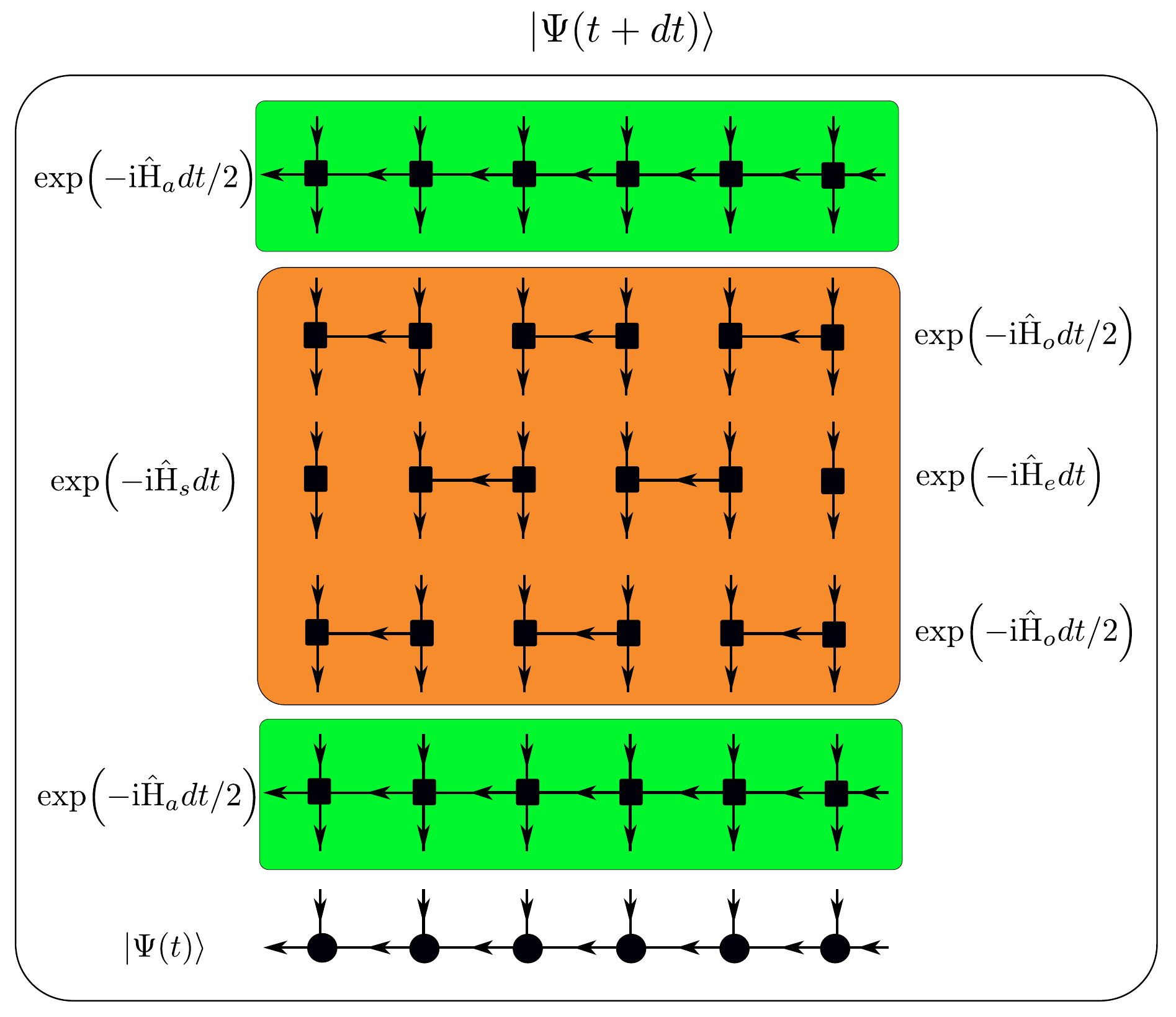}
\caption{Schematic representation of time evolution with $\DSMPS$s and $\DSMPO$s. Starting from $|\Psi(t)\rangle$ it is possible to compute $|\Psi(t+dt)\rangle$ doing a double Suzuki-Trotter decomposition. On one side the evolution is divided between the two contributions of the asymmetric, non-number conserving, Hamiltonian $\Hop_a$, each for a time $dt/2$ ($\DSMPO$s with green background), and the contribution from the symmetric part of the Hamiltonian $\Hop_s$ for a time $dt$ which here, for example, is performed with an ulterior Suzuki-Trotter decomposition between even ($\Hop_e$) and odd ($\Hop_o$) bonds. This part is represented over the orange background.} \label{fig:fig2} 
\end{figure}
Time evolution with $\DSMPS$ can be implemented in various ways. In the following we describe two of them. 

The first approach is based on multiplication between MPOs and MPSs, as in time evolution with Tchebychev polynomials \cite{Garcia-Ripoll2006, HalimehMcCulloch2015, ZaletelPollmann2014}. These type of algorithms only depend on MPO and MPS arithmetic, namely MPO multiplication with MPS and the addition and subtraction of MPSs. Hence these approaches can be readily implemented with $\DSMPS$. For a generic construction of MPO from arbitrary Hamiltonian (including symmetric ones), one can refer to \cite{ClaudiusSchollwock2017}. 

While the approach just described can be readily implemented, it is in general slower than Matrix Product States algorithms which use Suzuki-Trotter decomposition \cite{Trotter1959, Suzuki1976}. We thus describe here one way to use Suzuki-Trotter decomposition for $\DSMPS$. In these algorithms, one splits the time evolution operator $\Evo(dt)$ into many local operations which only affect the MPS locally. A typical example is shown in the central portion of Fig.\ref{fig:fig2}, with orange background, in which the evolution for a certain time $dt$ is divided in evolution on even and odd bonds (respectively with the Hamiltonians $\Hop_e$ and $\Hop_o$), and which can be parallelized (for more details on this $t-$MPS algorithm see, for instance, the review \cite{Schollwock2011}).    
In the following, for clarity, we focus on a unitary evolution with a Hamiltonian $\Ham$, however in the example discussed in Sec.\ref{ssec:dissi} we will show how this approach can also be readily applied to dissipative evolutions. 
For $\DSMPS$, special attentions should be paid to the terms of the Hamiltonian which change the total quantum number. Considering, for instance, that there is a local term $\hat{h}_l=\sgp_l + \sgm_l$ acting on the $l$-th site of an $\DSMPS$ initially in a single number sector $N$, the resulting state will be a superposition of three number sectors $N-1, N, N+1$. This means that even if $\hat{h}_l$ only acts locally on site $l$, the left boundary tensor at site $l=1$ of the $\DSMPS$ has to be updated and thus the effect on $\DSMPS$ is non-local. As a result, we should not simply absorb non-number conserving terms into a local operator and perform the usual Suzuki-Trotter based algorithms. 
One way to overcome this complication is to separate an Hamiltonian $\Ham$ into a symmetric, number-conserving part, $\Ham_s$, and an asymmetric, non-number-conserving part, $\Ham_a$:   
\begin{align}
\Ham = \Ham_s + \Ham_a.
\end{align}
It is now possible to use a second order time evolution operator as (setting $\hbar=1$)   
\begin{align}\label{eq:hybridtmps}
\Evo(\dt) = \exp\left(\frac{-\im\Ham_a\dt}{2}\right)\exp\left(-\im\Ham_s\dt\right)\exp\left(-\im\frac{\Ham_a\dt}{2}\right) 
\end{align}
as shown in Fig.\ref{fig:fig2}. 
For the symmetric portion, $\exp\left(-\im\Ham_s dt\right)$, one can perform a Suzuki-Trotter based algorithm, while for $\exp\left(-\im\Ham_a dt\right)$, one can treat it as an MPO and perform an MPO based time evolution. This hybrid time evolution algorithm would be efficient if the bond dimension $D_W$ of $\exp\left(-\im\Ham_a dt\right)$ is small which, for instance, is the case if the non-number-conserving term $\Ham_a$ is local. To give a more concrete idea before we discuss examples in the following section, local non-number conserving components of an Hamiltonian are typical in coupled resonator arrays, see for instance the review \cite{NohAngelakis2016}.

\section{Examples} \label{sec:app}

In the following we present two exemplary applications of $\DSMPS$. In Sec.\ref{ssec:hami} we discuss a unitary evolution while in Sec.\ref{ssec:dissi} we present a boundary driven dissipative systems. 

\subsection{Ground state and time evolution of an XYZ chain}\label{ssec:hami}
\begin{figure}
\includegraphics[width=\columnwidth]{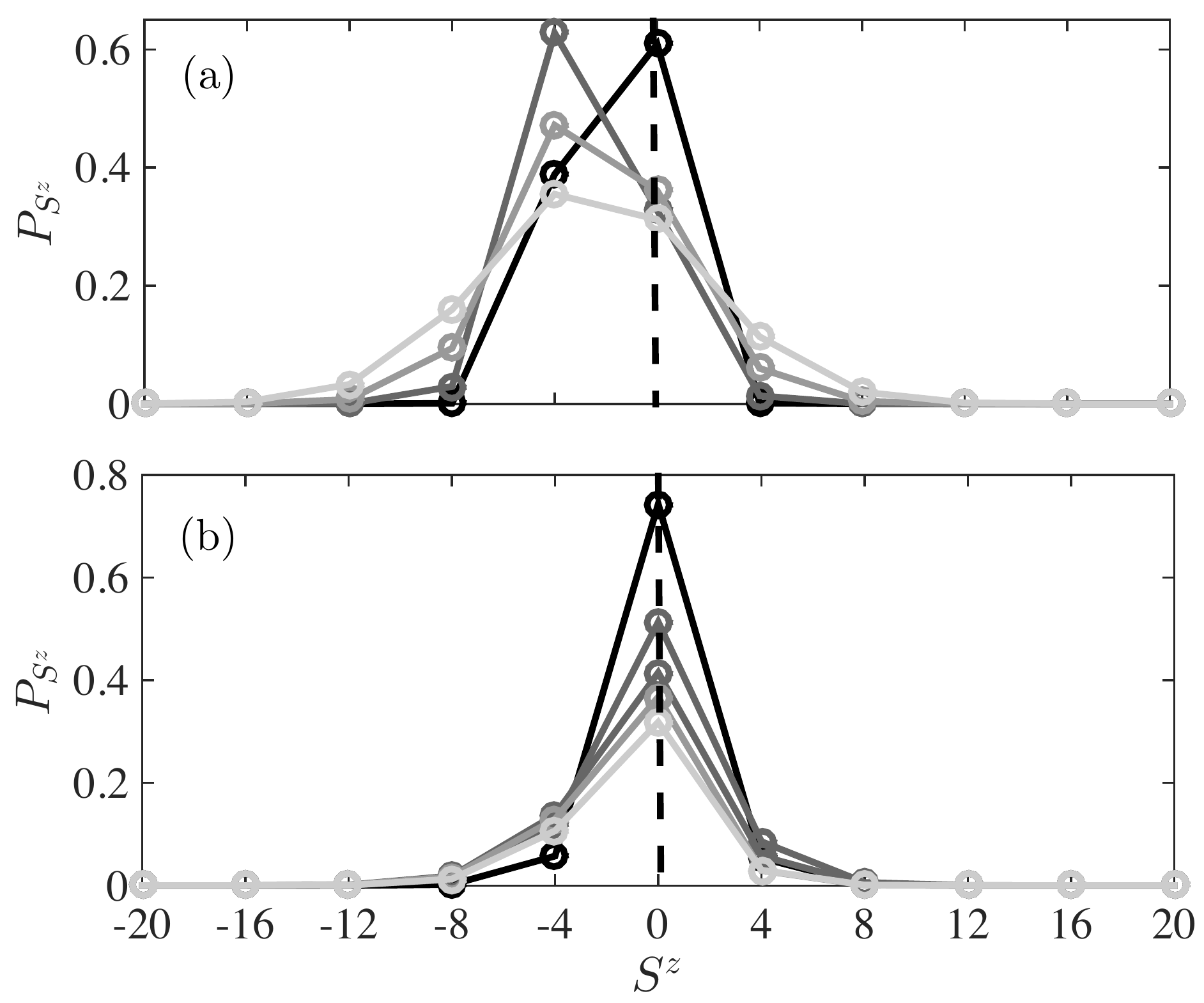}
\caption{(a) Distribution of ground states in different number sectors $P_{S^z_T}$. The lines from darker to lighter correspond $\gamma=0.1, 0.4, 0.7, 1$ respectively. The $\gamma=0$ case has a single total quantum number shown by the black dashed line. (b) Time evolution of the distribution of the quantum state in different number sectors, after a quench from $\gamma=0$ to $\gamma=0.5$. The initial state is the ground state of an XXZ chain with a fixed total spin as shown by the black dashed line. The lines from darker to lighter colors correspond to times $t=0.1, 0.2, 0.3, 0.4, 0.5$ respectively. The system size is $L=50$, the interaction strength $\Delta=1.5$ and the magnetization strength $h=0.5$. We have used bond dimensions up to $D=2000$. In both figures, only the data points with non-zero values are shown (indicated with circles).}  \label{fig:fig3} 
\end{figure}

The Hamiltonian of an $\XYZ$ spin chain of size $L$ can be written as
\begin{align}
\Hop_{{\rm XYZ}}(\gamma) = \sum_{l=1}^{L-1}&J_{{\rm XY}}\left[\left(1+\gamma\right)\sgx_l\sgx_{l+1} + \left(1-\gamma\right)\sgy_l\sgy_{l+1} \right. \nonumber \\ &+ \left. \Delta \sgz_l\sgz_{l+1} \right] + h\sum_{l=1}^L \sgz_l, \label{eq:XYZ}   
\end{align}
where $\sgx_l$, $\sgy_l$ and $\sgz_l$ are the operators corresponding to Pauli matrices, $J_{{\rm XY}}$ is the tunneling, $\Delta$ is the interaction strength and $\gamma$ denotes the anisotropy ranging from $0$ to $1$, $h$ is the strength of magnetization. For the special case $\gamma=0$, this model reduces to an $\XXZ$ chain with longitudinal field which conserves the total spin $\hat{S}^z_T=\sum_l \sgz_l$ and has $U(1)$ symmetry, while for $\gamma \neq 0$, the Hamiltonian commutes with the parity operator $\SYM{P}$ defined as
\begin{align}
\SYM{P} = \exp\left(\im \pi \hat{S}^z_T\right),
\end{align}
thus having a discrete $Z_2$ symmetry. In the following we work in units for which $J_{{\rm XY}}=\hbar=1$. We first apply the ground state search algorithm to $\Hop_{{\rm XYZ}}$ with $L=50$. In the following we keep the interaction strength $\Delta=1.5$ and magnetization strength $h=0.5$, and we compute the ground state for different values of $\gamma$, starting the search of the ground state from a state with zero total magnetization $\hat{S}^Z_T=0$. For the case $\gamma=0$, the resulting ground state will be the ground state in the same number sector as the trial starting state, namely in the number sector with $0$ total spin. However, for $\gamma\neq0$, the resulting ground state will automatically be in the even number sector, which is a superposition of states from different number sectors that all have an even number for the total spin. If one uses a $Z_2$ symmetric MPS for the $\gamma \neq 0$ case, one will obtain the same ground state but only in the number sector labelled by $0$ (corresponding to $\mod\!(\hat{S}^z_T,2)=0$ , i.e. total spin modulo $2$ is $0$). A major difference between using a $Z_2$ symmetric MPS and our $\DSMPS$ is that with our method it is straightforward to extract the information about the occupation of states with different total spins from the ground state. In fact, one can simply split an adaptively symmetric MPS into many $U(1)$ symmetric MPSs with a fixed total spin as in Eq.(\ref{eq:splitmps}), by just splitting the tensor on the left boundary. Therefore, with adaptively symmetric MPS, one can easily compute the distribution of the ground state in different number sectors $P_{S^z_T}$, which is shown in Fig.\ref{fig:fig3}(a). One can see clearly that for $\gamma=0$, the ground state has a single total spin (a kronecker delta at $0$ represented by a dashed line), while the distribution becomes broader as $\gamma$ increases to $1$. we can also see that the non-zero values are in a period of $4$, which means the total number of spins pointing up (down) is even.

We benckmark our $\DSMPS$ and $\DSMPO$ based ground state search algorithm with the conventional ground state search algorithm based on a $Z_2$ $\SMPS$ and $\SMPO$. For the parameters which we have considered in Fig.\ref{fig:fig3}(a), the differences between the ground state energies computed from both approaches are of the order $10^{-9}$, with a bond dimension $D=2000$. These two approaches use around $1.7$ and $2.5$ Gb memory, $0.75$ and $2$ hours separately on a personal computer with $4$ cores of $3.7$ GHz frequency.

We then do a quench, starting from the ground state corresponding to $\gamma=0$, $\vert GS_{{\rm XXZ}}\rangle$, and then evolving it with the Hamiltonian (\ref{eq:XYZ}) for $\gamma=0.5$, as if a sudden quench occurred. This evolution can be represented as
\begin{align}
\vert \psi(t) \rangle = \exp\left(-\im \Hop_{{\rm XYZ}}(\gamma=0.5) \; t \right) \vert GS_{{\rm XXZ}}\rangle. 
\end{align}
We chose to perform the evolution using a fourth order Runge-Kutta method based on MPO$-$MPS multiplication. 
To ensure convergence of our results, we have done simulations with bond dimensions $D=1000,1500,2000$ and checked various observables, for instance, for the average spin occupation $\langle \Psi(t)\vert \sgz_l\vert\Psi(t)\rangle$ we get a difference of the order of $10^{-9}$. We have also compared these observables with those computed with time-evolving matrix product states algorithm (t-MPS) \cite{Schollwock2011} using a $Z_2$ symmetric MPS and MPO, obtaining a difference of the order $10^{-4}$. We note that this difference is partially due to the fact that we have used a simple MPO based Runge-Kutta method for the time evolution, which is known to have poorer performance than other MPO based time evolution algorithms such as the approaches in \cite{ZaletelPollmann2014}. Instead, for the symmetric MPS with a $Z_2$ symmetry, we have used the standard t-MPS algorithm (by manually converting the initial state from $U(1)$ symmetry into $Z_2$ symmetry) which is known to be very stable and accurate. With the $\DSMPS$ approach it is straightforward to see that the quantum state of the system gradually expands into different number sectors with an even number of total spin $S^z_T$, as shown in Fig.\ref{fig:fig3}(b). We note that despite the initial state is computed for an Hamiltonian with a different symmetry from that of the evolution Hamiltonian, we use the same structure of Matrix Product States both for searching the ground state and the evolution, making the passage between the two different symmetry regimes seamless. 

\subsection{Time evolution of a dissipatively boundary driven Bose-Hubbard chain}\label{ssec:dissi} 

\begin{figure}
\includegraphics[width=\columnwidth]{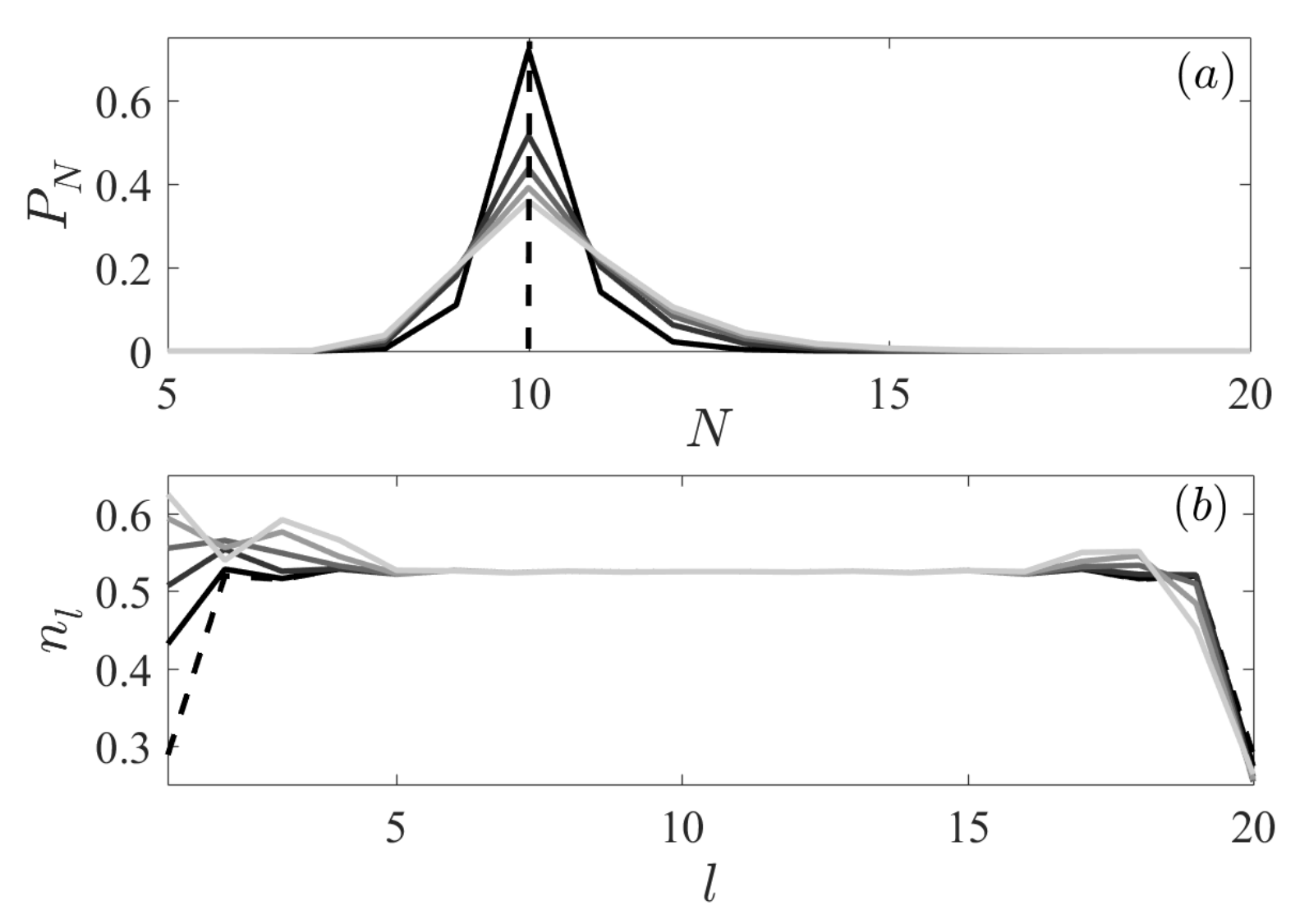}
\caption{(a) Time evolution of the distribution of the density operator in different number sectors, $P_N$ as in Eq.(\ref{eq:P_N}), starting from the ground state of a Bose-Hubbard chain with average filling $\bar{n}=0.5$. (b) Average occupation of each site $l$, $\langle n_l^{}\rangle$. In both panels, the black dashed line corresponds to the initial state. The lines from darker to lighter colors correspond to $t=0.2, 0.4, 0.6, 0.8, 1$ respectively. The system size $L=20$. The other parameters used are $d=10$, $U=4$, $\Lambda_1=\Lambda_L=1$ (in units for which $J=\hbar=1$), $\bar{n}_1=0.75$, $\bar{n}_{L}=0.25$ and the bond dimension $D=1500$. }  \label{fig:fig4} 
\end{figure}

The method to evolve in time the $\DSMPS$ can be readily extended from unitary to dissipative dynamics. We show an example in the following. The dynamics of the system is described by a master equation of Gorini-Kossakowski-Sudarshan-Lindblad form \cite{GoriniSudarshan1976, Lindblad1976} 
\begin{align} \label{eq:master}
\frac{d\rhoop}{dt} = \Lop(\rhoop) = -\frac{\im}{\hbar} [\Hop_{{\rm BH}}, \rhoop] + \Dop(\rhoop),
\end{align}
where we refer to $\Lop$ as the Lindbladian, $\Dop$ as dissipator, and where the Hamiltonian $\Hop_{{\rm BH}}$ can be written as
\begin{align} \label{eq:hami} 
\Hop_{{\rm BH}} =& -J\sum_{l=1}^{L-1}\left(\aop^{}_l\adop_{l+1} + \hc\right) \\ \nonumber 
&+ \frac{U}{2}\sum_{l=1}^L \nop^{}_l(\nop^{}_l-1).    
\end{align}
The dissipator can be written as
\begin{align} 
\Dop(\rhoop) = \sum_{l=1, L} &\Lambda_l \left[ \left(\bar{n}_l^{}+1\right)\left(2\aop_l^{}\rhoop\adop_l-\{\adop_l\aop_l^{}, \rhoop\}\right) \right. \nonumber \\ &+ \left. \bar{n}_l^{}\left(2\adop_l\rhoop\aop^{}_l-\{\aop^{}_l\adop_l, \rhoop\}\right) \right]. \label{eq:dissipator}    
\end{align}
In Eqs.(\ref{eq:hami},\ref{eq:dissipator}) we have used the notations $\adop_l, \aop_l^{}$ for the bosonic creation and annihilation operators and $\nop_l^{} = \adop_l\aop_l^{}$. $J$ is the tunneling amplitude and $U$ is the local interaction strength. $\Lambda_{1,L}$ denotes the strength of the dissipation acting at the edges $l=1$ or $l=L$. If $J=0$, the dissipator will impose a local thermal distribution at the edges with average occupation $\bar{n}_{1,L}^{}$ respectively. Similarly as for the case of the $\XYZ$ chain, in the following we work in units for which $J = \hbar=1$.

In general, to simulate the time evolution of this system with Matrix Product States, it is first useful to stress that, while at a given site it would be possible to have a very large (even infinite) number of bosonic particles, the first step is to limit the local Hilbert space size to a finite number, which we refer to as $d$. For unitary systems, for an interaction strength $U \geq 3$, it is usually sufficient to keep $d=4$. However, to faithfully represent a local bosonic thermal state, $d$ has to be much larger due to the long tail of the thermal distribution. In our simulations we have used $d=10$ and $d=11$ to ensure the accuracy of the results. 

A standard way to deal with density matrices with Matrix Product States is to consider them as vectors and, as a result, the local Hilbert space will be of size $d^2$. If one uses non-symmetric Matrix Product States for the time evolution, for instance splitting the operator $\exp\left(\Lop dt\right)$ into many two-sites operators, then each two-body operator would be a $d^4\times d^4$ matrix. For $d=10$, one such operator will contain $10^8$ (possibly complex) numbers, thus consuming a large amount memory. Furthermore, the most numeric expensive part of the Matrix Product States algorithm would be a two-site singular value decomposition performed on a $Dd^2\times Dd^2$ matrix. For a reasonable bond dimension $D=100$ and local Hilbert space $d=10$, this tensor would have a size of $10^4\times 10^4$, for which it is prohibitive to run a singular value decomposition on most personal computers. On this point we remark that in \cite{GuoPoletti2015} we kept only the diagonal and a few off-diagonal elements of the single-site reduce density matrix thus reducing the local space from $d^2$ to $k d$ with $k< 5$. This was possible for the dissipator used (same as the one in Eq.(\ref{eq:dissipator}), and the system analyzed), as the local density matrix is mainly diagonal. Such approximation is however not sufficient to study larger systems. Using $\DSMPS$ it is possible, as explained in the following, to study larger systems.     

We note that the unitary part of $\Lop$ is number conserving, while the dissipator in Eq.(\ref{eq:dissipator}) is local, thus the MPO for the non-number-conserving part has a bond dimension equal to $1$, which is the smallest possible size. In addition, the dissipator in Eq.(\ref{eq:dissipator}) has an ulterior property: if we write the density operator using the basis $|\vec{n}_N;N\rangle\langle \vec{m}_M;M|$ where $N$ is the total number of bosons in the ket ($M$ in the bra), and $\vec{n}_N$ is the vector detailing how the $N$ bosons are distributed over the $L$ sites (respectively $\vec{m}_M$ describes how the $M$ bosons are distributed between $L$ sites), then it appears clear that the dissipator couples the element corresponding to $|\vec{n}_N;N\rangle\langle \vec{m}_M;M|$ with $|\vec{n}_{N\pm 1};N\pm 1\rangle\langle \vec{m}_{M\pm 1};M\pm 1|$. This implies that if the initial condition belongs to a single number sector (e.g. at initial time $N=M$), then the density operator  will have only non-zero terms for blocks in which $N=M$. This block-diagonal structure of the density operator has been discussed in \cite{GuoPoletti2017b} and exploited in the context of studying transport with exact diagonalization tools. This symmetry of the density operator will automatically be preserved with $\DSMPS$ and $\DSMPO$, making our hybrid time evolution algorithm introduced in Sec.\ref{ssec:timeevolution} an ideal tool to study this type of problems. 

In our exemplary simulation, we first prepared the initial state of the system to be the ground state of a Bose-Hubbard chain of size $L=20$ with average filling $0.5$ and $U=4$, which we denote as $\vert GS_{{\rm BH}} \rangle$. We then turn on the dissipation, with $\Lambda_{1} = \Lambda_L=1$, $\bar{n}_1 = 0.75$ and $\bar{n}_L = 0.25$. 
For the simulations we have truncated the local Hilbert space to $d=10$ and used a bond dimension $D=1000$ for the MPS, and the time step $dt=0.01$. 
To ensure that our choices of $d$ and $D$ are appropriate, we have also done simulations and computed the local density $\langle \hat{n}_j\rangle=\tr\left(\hat{n}_l \; \rhoop\right)$ with $d=11$, getting a difference of the order of $10^{-5}$, and with $D=1500$, getting a difference of the order of $10^{-4}$.    
In Fig.\ref{fig:fig4}(a), we plot the distribution of the quantum state in different number sectors, i.e. 
\begin{align} 
P_N=\sum_{\vec{n}_N} \langle \vec{n}_N ; N | \rhoop |\vec{n}_N ; N \rangle, \label{eq:P_N} 
\end{align} 
something readily done with $\DSMPS$. The initial state has a total of $10$ bosons, then during the evolution, the state becomes a superposition of states from many number sectors due to the boundary dissipative driving. The distribution of the density operator in different number sectors becomes broader with time. In Fig.\ref{fig:fig4}(b), we plot the average occupation $\langle \hat{n}_l\rangle$ as a function of site $l$ at different times. Initially, since the evolution starts from the ground state, the distribution of $\langle n_l\rangle$ is symmetric around the middle site. At later times the distribution becomes unbalanced due to the different drivings at the two edges.

\section{conclusion} \label{sec:summary}
In this work we have proposed a Matrix Product States algorithm that can treat a global $U(1)$ symmetry or one of its subgroups on the same footing. At the same time, this method can also deal with non-number conserving systems. We have shown how this method, based on adaptively symmetric Matrix Product States, can be applied to search for the ground state of a system and also for time evolution. 

We shall note that the conventional symmetric Matrix Product States method can be viewed as a special case of the one presented here. Moreover, for systems without $U(1)$ symmetry, or, for instance, with only $Z_2$ symmetry, the adaptively symmetric Matrix Product States method allows to readily acquire additional information about the distribution of the state in different number sectors. 

This method could be very useful in some applications in which the presence of symmetries, and their type, changes within the evolution. We have studied two such examples both for unitary and dissipative systems. In both cases the use of adaptively symmetric Matrix Product States allows to readily follow the evolution of the system and benefit of the symmetries when present. The efficiency of the method depends on the system studied.  
Adaptively symmetric Matrix Product States could also be extended to systems with non-Abelian symmetries, however this is beyond the scope of the current work.

\begin{acknowledgments} 
C. G. acknowledges support from National Natural Science Foundation of China under Grants No. 11504430 and No. 11805279. D.P. acknowledges support from the Singapore Ministry of Education, Singapore Academic Research Fund Tier-II (project MOE2016-T2-1-065). 
\end{acknowledgments}

\appendix

\section{Construction of $\DSMPS$ and $\DSMPO$ for XYZ chain}

In this appendix we demonstrate the process of constructing of an $\DSMPS$ and an $\DSMPO$ using the concrete example of a $3$-site spin XYZ chain with the magnetization strength $h=0$. We label the state $\vert0\rangle$ with quantum number $0$, and the state $\vert1\rangle$ with $1$. Assuming we have an initial state which is 

\begin{align}
\vert\psi\rangle = \frac{\sqrt{2}}{2}\left(\vert 110 \rangle + \vert 100\rangle\right)
\end{align}

The $\DSMPS$ corresponding to state $\vert \psi \rangle$ is 

\begin{align}
\SYM{M}_1 &= \frac{\sqrt{2}}{2} \{\textbf{1}^{\cev{1}}_{\vec{1}, \cev{0}}, \textbf{1}^{\cev{1}}_{\vec{2}, \cev{1}} \}, \nonumber \\
\SYM{M}_2 &= \{\textbf{1}^{\cev{0}}_{\vec{0}, \cev{0}}, \textbf{1}^{\cev{1}}_{\vec{1}, \cev{0}} \} , \nonumber \\ 
\SYM{M}_3 &= \{\textbf{1}^{\cev{0}}_{\vec{0}, \cev{0}} \}.  
\end{align}
In fact on site $l=3$ there is only one possible quantum number flowing in $\cev{a}_4=0$, only one quantum number present on the site $\cev{\sigma}_3=0$ and hence only one possible output $\vec{a}_3=0$. Similarly, at site $l=2$ there are two possible values for $\cev{\sigma}_2=0,\;1$ which results in $\vec{a}_	2=0,\;1$. At site $l=3$, the local quantum number is only $\cev{\sigma}_1=1$, but since $\cev{a}_2$ takes two possible values, then $\vec{a}_	1=1,\;2$. 

In the non-symmetric case, the MPO corresponding to $\Hop_{{\rm XYZ}}$ for $L=3$ can be straightforwardly written as

\begin{align}
\MPO_{\rm XYZ} = \left[\begin{array}{cccc}
(1+\gamma)\sigma^x & (1-\gamma) \sigma^y & \Delta \sigma^z & \id
\end{array}\right] \otimes \nonumber \\ 
\left[\begin{array}{cccc}
\sigma^x & \zero & \zero & \zero \\
\sigma^y & \zero & \zero & \zero \\
\sigma^z & \zero & \zero & \zero \\
\id & (1+\gamma)\sigma^x & (1-\gamma) \sigma^y & \Delta \sigma^z \\
\end{array}\right] \otimes \left[\begin{array}{c}
\id \\
\sigma^x  \\
\sigma^y  \\
\sigma^z  \\
\end{array}\right].
\end{align}
Now we first rewrite $\Hop_{{\rm XYZ}}$ in the following form
\begin{align}
\Hop_{{\rm XYZ}} = 2\sum_{j=1}^2 &\left[\sgp_j\sgm_{j+1}+\sgm_j\sgp_{j+1}+ \right. \nonumber \\ & \left. \gamma(\sgp_j\sgp_{j+1} + \sgm_j\sgm_{j+1}) + \Delta\sgz_j\sgz_{j+1}\right],
\end{align}
which is a summation of $10$ terms (because there are two bonds). To write down the corresponding $\DSMPO_{\rm XYZ}$, it is convenient to first write the product MPO for each term and then use the MPO addition rule to sum them up \cite{ClaudiusSchollwock2017}. The product MPO corresponding to the terms $\sgp_1\sgm_2$, $\sgm_1\sgp_2$, $\sgp_1\sgp_2$, $\sgm_1\sgm_2$, $\sgz_1\sgz_2$ can be written as

\begin{align}
\MPO_{\sgp_1\sgm_2} =&\{\id_{\vec{0}, \cev{-1}}^{\vec{0}, \cev{1}}\} \times \{\id_{\vec{-1}, \cev{0}}^{\vec{1}, \cev{0}}\} \times \{\id_{\vec{0}, \cev{0}}^{\vec{0}, \cev{0}}, \id_{\vec{0}, \cev{0}}^{\vec{1}, \cev{1}} \} \\ 
\MPO_{\sgm_1\sgp_2} =&\{\id_{\vec{0}, \cev{1}}^{\vec{1}, \cev{0}}\} \times \{\id_{\vec{1}, \cev{0}}^{\vec{0}, \cev{1}}\} \times \{\id_{\vec{0}, \cev{0}}^{\vec{0}, \cev{0}}, \id_{\vec{0}, \cev{0}}^{\vec{1}, \cev{1}} \} \\
\MPO_{\sgp_1\sgp_2} =&\{\id_{\vec{2}, \cev{1}}^{\vec{0}, \cev{1}}\} \times \{\id_{\vec{1}, \cev{0}}^{\vec{0}, \cev{1}}\} \times \{\id_{\vec{0}, \cev{0}}^{\vec{0}, \cev{0}}, \id_{\vec{0}, \cev{0}}^{\vec{1}, \cev{1}} \} \\
\MPO_{\sgm_1\sgm_2} =&\{\id^{\vec{1}, \cev{0}}_{\vec{-2}, \cev{-1}}\} \times  \{\id_{\vec{-1}, \cev{0}}^{\vec{1}, \cev{0}}\} \times \{\id_{\vec{0}, \cev{0}}^{\vec{0}, \cev{0}}, \id_{\vec{0}, \cev{0}}^{\vec{1}, \cev{1}} \} \\
\MPO_{\sgz_1\sgz_2} =&\{\id_{\vec{0}, \cev{0}}^{\vec{0}, \cev{0}}, -\id_{\vec{0}, \cev{0}}^{\vec{1}, \cev{1}} \} \times \nonumber \\
 &\{\id_{\vec{0}, \cev{0}}^{\vec{0}, \cev{0}}, -\id_{\vec{0}, \cev{0}}^{\vec{1}, \cev{1}} \} \times \{\id_{\vec{0}, \cev{0}}^{\vec{0}, \cev{0}}, \id_{\vec{0}, \cev{0}}^{\vec{1}, \cev{1}} \} 
\end{align}

We notice that for the terms $\MPO_{\sgp_1\sgm_2}$, $\MPO_{\sgm_1\sgp_2}$, $\MPO_{\sgz_1\sgz_2}$, the auxiliary index on the left boundary ${b_1}=0$, while we have ${b_1}=2$ for the term $\MPO_{\sgp_1\sgp_2}$, and ${b_1}=-2$ for the term $\MPO_{\sgm_1\sgm_2}$. We can also construct the remaining terms of $\MPO_{\sgp_2\sgm_3}$, $\MPO_{\sgm_2\sgp_3}$, $\MPO_{\sgp_2\sgp_3}$, $\MPO_{\sgm_2\sgm_3}$, $\MPO_{\sgz_2\sgz_3}$ similarly. We note that if  we encode the MPO with $Z_2$ symmetry instead of $U(1)$ symmetry, then ${b_1}$ for the terms $\MPO_{\sgp_1\sgp_2}$ and $\MPO_{\sgm_1\sgm_2}$ will also be $0$ since $2$ and $-2$ are equivalent to $0$ modulo $2$. The final $\DSMPO$ is a summation of all these terms

\begin{align}
\DSMPO_{\XYZ} = 2\sum_{j=1}^2 &\left[\MPO_{\sgp_j\sgm_{j+1}}+\MPO_{\sgm_j\sgp_{j+1}} \right. \nonumber \\ &+ \left. \gamma(\MPO_{\sgp_j\sgp_{j+1}} + \MPO_{\sgm_j\sgm_{j+1}}) \right. \nonumber \\ &+ \left. \Delta\;\MPO_{\sgz_j\sgz_{j+1}}\right].
\end{align}


\begin{thebibliography}{99}
\bibitem{FannesWerner1992} M. Fannes, B. Nachtergaele, and R. Werner, Commun. Math. Phys., {\bf 144}, 443 (1992).
\bibitem{OstlundRommer1995} S. Ostlund and S. Rommer, Phys. Rev. Lett., {\bf 75}, 3537 (1995). 
\bibitem{Perez-GarciaCirac2007} D. Perez-Garcia, and F. Verstraete, M.M. Wolf, and J.I. Cirac, Quantum Inf. Comput., {\bf 7}, 401 (2007). 
\bibitem{Schollwock2011} U. Schollwo\"ck, Ann. Phys. {\bf 326}, 96 (2011).

\bibitem{Hastings2007} M. B. Hastings, J. Stat. Mech: Theory Exp. (2007) {\bf P}08024.
\bibitem{Vidal2008} G. Vidal, Phys. Rev. Lett., {\bf 101}, 110501 (2008).


\bibitem{Vidal2003} G. Vidal, Phys. Rev. Lett. {\bf 91}, 147902 (2003).   
\bibitem{Vidal2004} G. Vidal, Phys. Rev. Lett. {\bf 93}, 040502 (2004).         
\bibitem{DaleyVidal2004} A. J. Daley, C. Kollath, U. Schollw\"ock, and G. Vidal, J. Stat. Mech. Theor. Exp., (2004) {\bf P}04005.
\bibitem{WhiteFeiguin2004} S. R. White and A. E. Feiguin, Phys. Rev. Lett. {\bf 93}, 076401 (2004). 
\bibitem{VerstraeteCirac2004} F. Verstraete, J. J. Garc\'ia-Ripoll, and J. I. Cirac, Phys. Rev. Lett., {\bf 93}, 207204 (2004).

\bibitem{Daley2014} A. J. Daley, Adv. Phys. {\bf 63}, 77 (2014)  
\bibitem{DeVegaBanuls2015} I. de Vega, and M.C. Banuls, Phys. Rev. A (2015)   
\bibitem{GuoPoletti2018a} C. Guo, I. de Vega, U. Schollw\"ock, and D. Poletti, Phys. Rev. A {\bf 97}, 053610 (2018). 
\bibitem{XuPoletti2019} X. Xu, J. Thingna, C. Guo, and D. Poletti, Phys. Rev. A {\bf 99}, 012106 (2019). 
\bibitem{ProsenZnidaric2009} T. Prosen, and M. Znidaric, J. Stat. Mech. (2009) {\bf P}02035. 
\bibitem{PalmeroPoletti2019} M. Palmero, X. Xu, C. Guo, and D. Poletti, arXiv:1901.05145 (2019). 



\bibitem{PizornProsen2008} T. Prosen, and I. Pizorn, Phys. Rev. Lett. {\bf 101}, 105701 (2008). 


\bibitem{Perez-GarciaVidal2008} D. Perez-Garcia, M.M. Wolf, M. Sanz, F. Verstraete, and J.I. Cirac, Phys. Rev. Lett., {\bf 100}, 167202 (2008).
\bibitem{SinghVidal2011} S. Singh, R.N.C. Pfeifer, and G. Vidal, Phys. Rev. B {\bf 83}, 115125 (2011).
\bibitem{McCulloch2007} I. P. McCulloch, J. Stat. Mech. (2007) {\bf P}10014.


\bibitem{Garcia-Ripoll2006} J. J. Garc\'ia-Ripoll, New J. Phys. {\bf 8}, 305 (2006).
\bibitem{HalimehMcCulloch2015} J.C. Halimeh, F. Kolley, and I.P. McCulloch, Phys. Rev. B {\bf 92}, 115130 (2015).
\bibitem{ZaletelPollmann2014} M.P. Zaletel, R.S.K. Mong, C. Karrasch, J.E. Moore, and F. Pollmann, Phys. Rev. B {\bf 91}, 165112 (2015).

\bibitem{ClaudiusSchollwock2017} C. Hubig, I. P. McCulloch, and U. Schollwo\"ck, Phys. Rev. B {\bf 95}, 035129 (2017).

\bibitem{Trotter1959} H.F. Trotter, Proc. Amer. Math. Soc. {\bf 10}, 545 (1959).  
\bibitem{Suzuki1976} M. Suzuki, Comm. Math. Phys. {\bf 51}, 2 (1976).

\bibitem{NohAngelakis2016} C. Noh, and D. Angelakis, Rep. Prog. Phys. {\bf 80}, 016401 (2016).  


\bibitem{GoriniSudarshan1976} V. Gorini, A. Kossakowski, and E. C. G. Sudarshan, J. Math. Phys. {\bf 17}, 821 (1976).
\bibitem{Lindblad1976} G. Lindblad, Commun. Math. Phys. {\bf 48}, 119 (1976).

\bibitem{GuoPoletti2015} C. Guo, M. Mukherjee, and D. Poletti, Phys. Rev. A {\bf 92}, 023637 (2015). 

\bibitem{GuoPoletti2017b} C. Guo, and D. Poletti, Phys. Rev. B {\bf 96}, 165409 (2017).









\end{thebibliography}
\end{document}